\documentclass[sigconf,natbib=true,anonymous=false]{acmart}
\usepackage{booktabs}
\usepackage{multirow}
\usepackage{subcaption}
\usepackage{graphicx}
\usepackage{footnote}
\usepackage{tablefootnote}
\AtBeginDocument{%
  \providecommand\BibTeX{{%
    \normalfont B\kern-0.5em{\scshape i\kern-0.25em b}\kern-0.8em\TeX}}}

\begin{document}

\title{IA-GCN: Interactive Graph Convolutional Network for Recommendation}

% \author{Yinan Zhang}
\author{Yinan Zhang*, Pei Wang*, Congcong Liu}
\author{Xiwei Zhao, Hao Qi, Jie He, Junsheng Jin,
Changping Peng, Zhangang Lin, Jingping Shao}
\email{yinan.zhang1996@gmail.com,wangpei102595@gmail.com,cliubh@connect.ust.hk} \email{{zhaoxiwei,qihao1,hejie67,jinjunsheng1,
pengchangping,linzhangang,shaojingping}@jd.com}
\affiliation{%
  \institution{Business Growth BU, JD.com}
  \city{Beijing}
  \country{China}
}
% \author{Pei Wang}
% \email{wangpei959@jd.com}
% \affiliation{%
%   \institution{Business Growth BU, JD}
%   \city{Beijing}
%   \country{China}
% }
% \author{Xiwei Zhao}
% \email{zhaoxiwei@jd.com}
% \affiliation{%
%   \institution{Business Growth BU, JD}
%   \city{Beijing}
%   \country{China}
% }
% \author{Hao Qi}
% \email{qihao1@jd.com}
% \affiliation{%
%   \institution{Business Growth BU, JD}
%   \city{Beijing}
%   \country{China}
% }
% \author{Jie He}
% \email{hejie67@jd.com}
% \affiliation{%
%   \institution{Business Growth BU, JD}
%   \city{Beijing}
%   \country{China}
% }
% \author{Junsheng Jin}
% \email{jinjunsheng1@jd.com}
% \affiliation{%
%   \institution{Business Growth BU, JD}
%   \city{Beijing}
%   \country{China}
% }
% \author{Changping Peng}
% \email{pengchangping@jd.com}
% \affiliation{%
%   \institution{Business Growth BU, JD}
%   \city{Beijing}
%   \country{China}
% % }
% \author{Zhangang Lin}
% \email{linzhangang@jd.com}
% \affiliation{%
%   \institution{Business Growth BU, JD}
%   \city{Beijing}
%   \country{China}
% }
% \author{Jingping Shao}
% \email{shaojingping@jd.com}
% \affiliation{%
%   \institution{Business Growth BU, JD}
%   \city{Beijing}
%   \country{China}

\begin{abstract}
\renewcommand*{\thefootnote}{\fnsymbol{footnote}}
\footnotetext{*This work was fulfilled when Yinan Zhang and Pei Wang worked at JD.COM.}
Recently, Graph Convolutional Network (GCN) has become a novel state-of-art for Collaborative Filtering (CF) based Recommender Systems (RS). It is a common practice to learn informative user and item representations by performing embedding propagation on a user-item bipartite graph, and then provide the users with personalized item suggestions based on the representations. Despite effectiveness, existing algorithms neglect precious interactive features between user-item pairs in the embedding process. When predicting a user's preference for different items, they still aggregate the user tree in the same way, without emphasizing target-related information in the user neighborhood. Such a uniform aggregation scheme easily leads to suboptimal user and item representations, limiting the model expressiveness to some extent.

In this work\footnotetext{This is the author’s original manuscript which is posted only for your personal use, not for redistribution. Another precise accepted manuscript was published in CIKM '23: Proceedings of the 32nd ACM International Conference on Information and Knowledge Management, October 2023, Pages 4410–4414, https://doi.org/10.1145/3583780.3615232.}, we address this problem by building bilateral interactive guidance between each user-item pair and proposing a new model named IA-GCN (short for InterActive GCN). Specifically, when learning the user representation from its neighborhood, we assign higher attention weights to those neighbors similar to the target item. 
Correspondingly, when learning the item representation, we pay more attention to those neighbors resembling the target user. 
This leads to interactive and interpretable features, effectively distilling target-specific information through each graph convolutional operation. Our model is built on top of LightGCN, a state-of-the-art GCN model for CF, and can be combined with various GCN-based CF architectures in an end-to-end fashion. Extensive experiments on three benchmark datasets demonstrate the effectiveness and robustness of IA-GCN. Codes are available at https://github.com/jd-opensource/IA-GCN.

% We conduct extensive experiments on three benchmark datasets, and IA-GCN consistently outperforms several state-of-the-art models like NGCF\cite{he2017neural}, LightGCN\cite{he2020lightgcn} and DGCF\cite{wang2020disentangled}. Further studies justify the rationality of our design and offer insights into the interpretability of attention weights in user-item pairs.

\end{abstract}




\maketitle

\section{Introduction}
% \begin{figure}[t]\begin{center}
% \includegraphics[width=8.4 cm]{graph1.pdf}\end{center}
% \caption{The two-tree structure 
% in GCN-based recommendation framework, $K=3$. 
% To predict the preference score of the target  $u-i$ pair from their $0-3$ order features,
% GCN would aggregate information within $u$/$i$'s $3$-hop neighborhood. Hence both trees consist of $3$ levels (excluding roots).
% Before the convolution, $\mathbf e^0$ of all nodes are initialized. Then
% Eq (\ref{eq:agg_general}) is used iteratively from bottom to top:
% aggregating $\mathbf e^0$ in level $3$ to get $\mathbf e^1$ in level $2$, aggregating $[\mathbf e^0,\mathbf e^1]$ in level $2$ to get $[\mathbf e^1,\mathbf e^2]$ in level $1$, and finally
% aggregating $[\mathbf e^0,\mathbf e^1,\mathbf e^2]$ in level $1$ to get $[\mathbf e^1, \mathbf e^2,\mathbf e^3]$ in level 0, for $u$ and $i$.  
% }\label{arch}
% \end{figure}
\begin{figure*}
\centering
    \begin{minipage}[b]{0.49\textwidth}
    \centering
    \includegraphics[width=\textwidth]{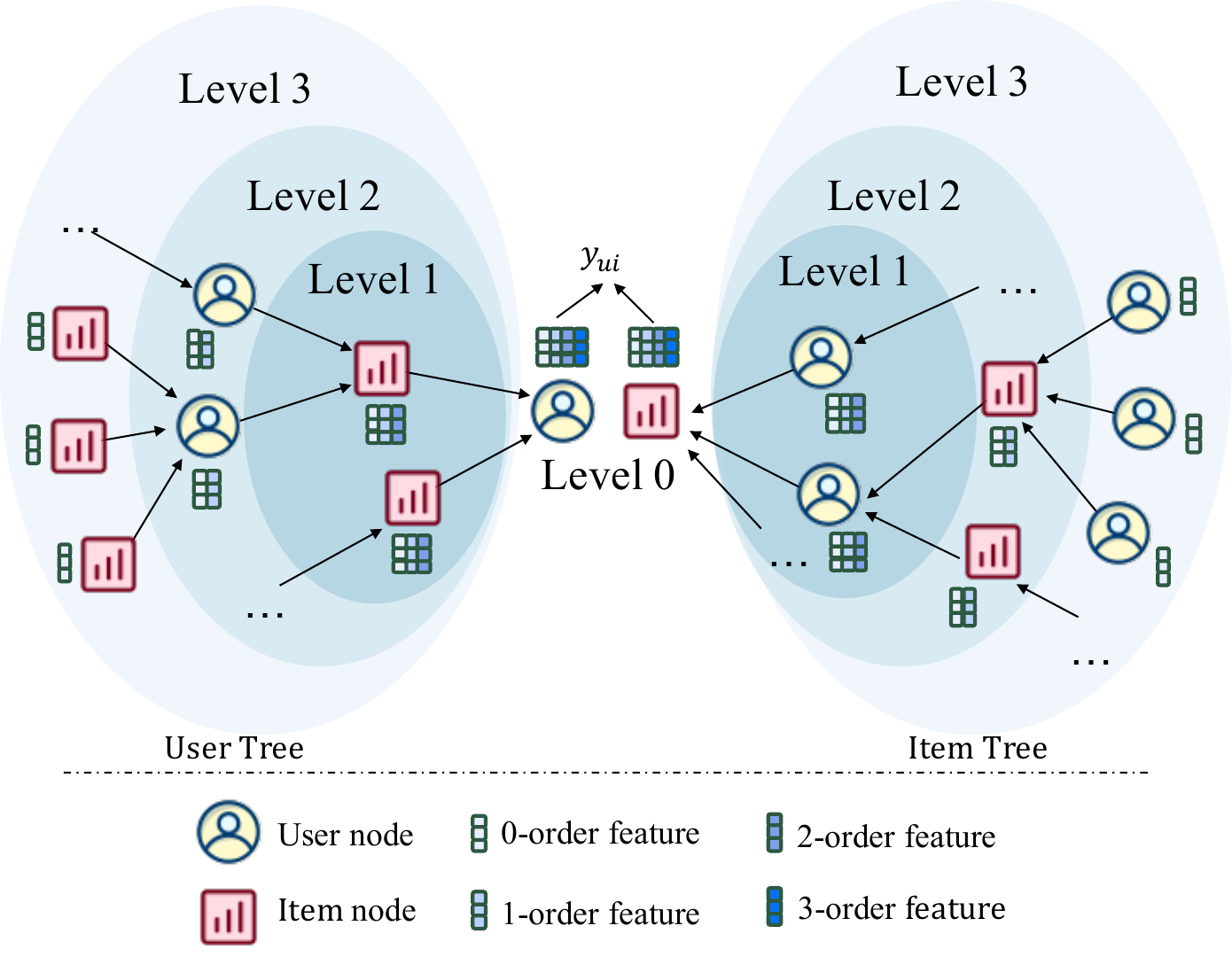}
    \subcaption{The two-tree structure in GCN-based recommendation framework}
    \label{arch}
    \end{minipage}
\hfill\vline\hfill
    \begin{minipage}[b]{0.49\textwidth}
    \centering
    \includegraphics[width=\textwidth]{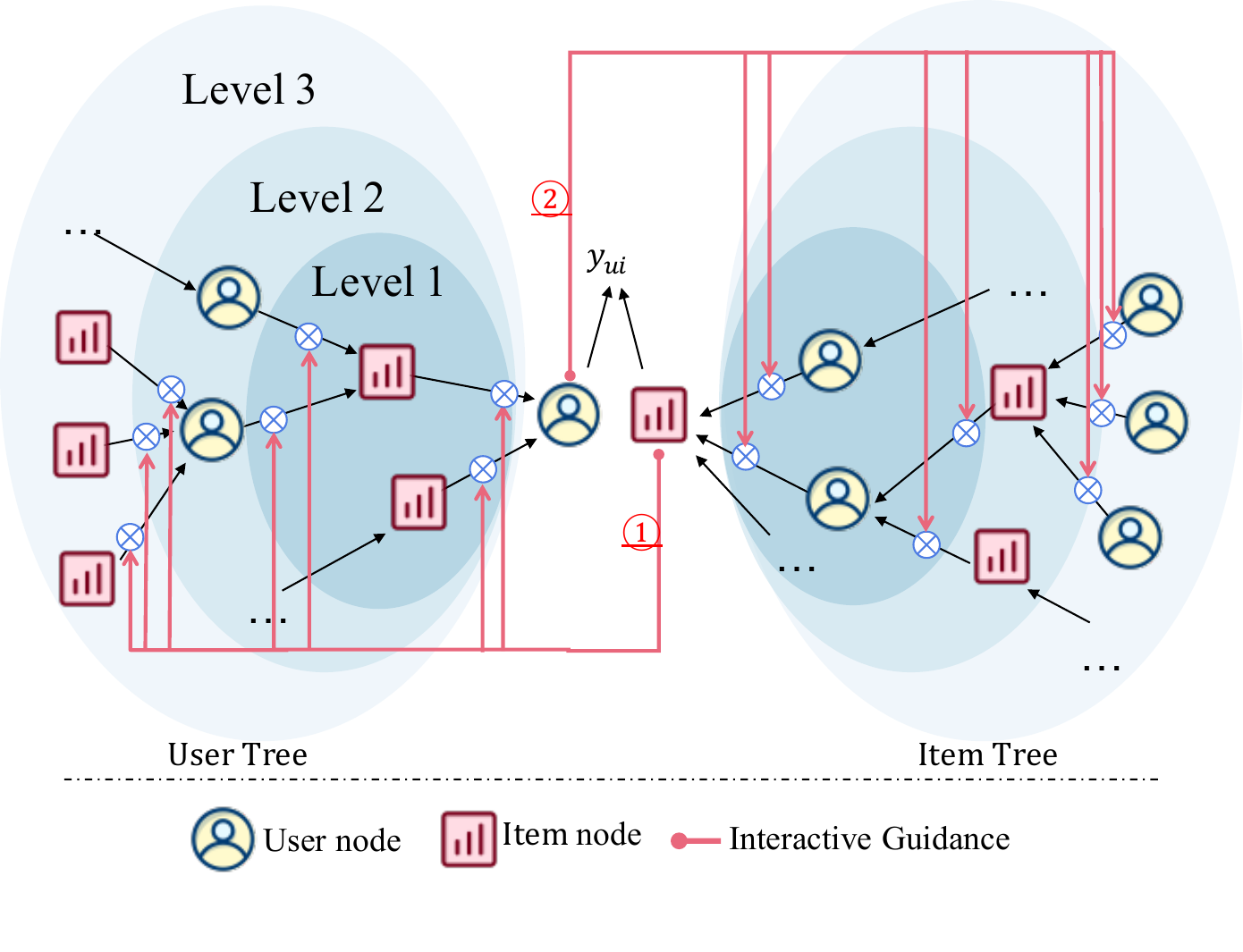}
        \subcaption{The proposed InterActive GCN structure.}
        \label{arch_inter}
    \end{minipage}
    % \centering
  \caption{\centering Illustration on GCN-based frameworks.}
  \medskip
\small
\flushleft
\textbf{(a)}: To predict the preference score of the target $u-i$ pair from their $0-3$ order features, GCN would aggregate information within $u$/$i$'s $3$-hop neighborhood. Hence both trees consist of $3$ levels (excluding roots). Before the convolution, $\mathbf e^0$ of all nodes are initialized. Then Eq.\eqref{eq:agg_general} is used iteratively from bottom to top: aggregating $\mathbf e^0$ in level $3$ to get $\mathbf e^1$ in level $2$, aggregating $[\mathbf e^0,\mathbf e^1]$ in level $2$ to get $[\mathbf e^1,\mathbf e^2]$ in level $1$, and finally aggregating $[\mathbf e^0,\mathbf e^1,\mathbf e^2]$ in level $1$ to get $[\mathbf e^1, \mathbf e^2,\mathbf e^3]$ in level 0, for $u$ and $i$.\\
  \textbf{(b)}: Different from existing GCN-based recommendation algorithms that late-fuse the $u-i$ features, IA-GCN learns the interactions between the two trees through an interactive guidance strategy. Specifically, each aggregator in the user /item tree is guided by the root to emphasize children that are similar to the target $i$/$u$. So the resulting high-order features contain precious target-specific features that will further strengthen the preference prediction.
  \label{fig:teaser}
\end{figure*}

In the era of the information explosion,
\textit{Recommender System} (RS) which helps us to filter out tremendous uninformative messages and reach interested ones,
plays a central role in many online services, ranging from e-commerce, advertising, social media to news outlets \cite{schafer2001commerce, tang2013social}.
Behind this, the core task of RS is to make predictions on a fundamental question: how likely a target user $u$ would interact with (click, purchase, rate, etc.) a target item $i$.
\textit{Collaborative Filtering} (CF) \cite{shi2014collaborative} successfully addresses this preference prediction problem by exploiting a large volume of historical user-item interactions, which makes it
almost the default framework in many real-world RSs. 

In general, the preference score of CF is predicted from the fusion (inner product \cite{koren2009matrix}, MLP \cite{he2017neural}, euclidean distance \cite{hsieh2017collaborative}, etc.) of two \textit{embedding} vectors that represent the latent features of the target user $u$ and the target item $i$ respectively.
As a result, how to build expressive embeddings to capture satisfactory user/item portraits is of crucial importance to the prediction performance.
Early CF algorithms, such as matrix factorization (MF), mostly directly project the user/item ID into an embedding vector \cite{koren2009matrix}.
Later, many enhance the target user $u$'s embeddings by considering $u$'s historical interactions as her pre-existing features in the embedding calculation \cite{zhou2019deep, liu2020kalman}. 
%Representative works include SVD++ \cite{}, NAIS \cite{}, ACF \cite{}, and also DIN \cite{},DSIN \cite{}, KFAtt \cite{} in industrial RSs.
%Taking advantage of $u$'s one-hop connectivity, these algorithms also inspire a novel insight: can we also make use of $u$/$i$'s higher-hop connectivity to further lift the expressiveness of $u$/$i$'s feature vectors?
Recent years have witnessed many emerging studies in \textit{Graph Convolutional Neural Network} (GCN) based CF algorithms, which further lift the expressiveness of $u$/$i$'s embedding vectors by exploiting the high-hop connectivity among users and items.
Representative works include Pinsage \cite{ying2018graph}, NGCF \cite{wang2019neural}, Light-GCN \cite{he2020lightgcn}, and among others \cite{wang2020disentangled, wu2021self}.
Specifically, the data structure of CF is naturally in a bipartite graph: 
users and items as nodes and interactions as edges. And the $K$-order features of node $u$/$i$, which summarize the information within $u$/$i$'s $K$-hop neighborhood, are aggregated by of $K$ stacked graph convolutional layers, forming a tree-like structure, the  \textit{user/item  tree}.
We illustrated this commonly used 
\textit{two-tree} structure in Fig.\ref{arch}.

Despite being extensively studied, existing GCN-based CF algorithms mostly suffer from a key limitation, no \textit{interaction} between the user tree and item tree until the final fusion in the CF layer.
This is because their aggregation is mostly inherited from conventional GCNs, e.g., GraphSage \cite{hamilton2017inductive}, that was originally proposed
for classification on every single node.
However, the recommendation task is fundamentally different from classification: it is not $u$ and $i$'s general portraits, e.g.,
$u$'s purchasing power and $i$'s ratings, but their interactive features, e.g., 
$u$'s consideration when choosing $i$ and $i$'s partial characteristic which attracts $u$, that determine the preference.
Suffering from this suboptimal late fusion architecture, existing algorithms miss the precious interactive features and are thus ineffective in the preference prediction.

To tackle this limitation, we propose InterActive GCN, a novel architecture specially designed to model the user-item interactions in GCN-based CF. Different from conventional GCNs that extract $u$ and $i$'s general features independently, IA-GCN builds explicit guidance links between the two trees (Fig.\ref{arch_inter}).
For the aggregations in the user tree, we allocate high importance to neighbors similar to the target item $i$. And correspondingly in the item tree, we emphasize neighbors similar to the target user $u$.
This interactive guidance enables GCN to focus on target specific information through every convolution and thus
capture interactive characteristics in $u$/$i$'s high-order features, which finally contributes to the significant performance gain in the preference prediction.
In summary, we make the following contributions:
\begin{itemize}
\item To the best of our knowledge, we are the first to highlight the negative impact of late fusion in aggregating $u$ and $i$'s high-order features in conventional GCN-based CF algorithms.
\item We propose IA-GCN, a novel
GCN architecture specifically
for CF. The key idea is to extract interactive features for $u$ and $i$ by building interactive guidance between the two trees which emphasizes target-specific information through each convolutional operation.
\item We validate the effectiveness of IA-GCN 
through extensive experiments on commonly used benchmark datasets.
IA-GCN outperforms a variety of state-of-the-art GCN-based CF algorithms, validating the significance of capturing interactive features.
\end{itemize}

\section{Related Works}
Our work is closely related to three active research areas: collaborative filtering, GCN for the recommendation, and attention mechanism.

\subsection{Collaborative Filtering}
% embedding and calculation
% The core of personalized recommendation is to predict the likelihood that a user will interact with an item, e.g., by clicking on it or making a purchase. 
Collaborative filtering (CF) assumes that users with similar preferences might also be interested in similar items \cite{shi2014collaborative}. 
A common paradigm of CF is to represent users and items via latent vectors and then attempt to derive probabilities for interactions by fusing the two embedding vectors. 
Matrix factorization embeds users and items into embeddings and directly uses the inner product to make predictions\cite{koren2009matrix}. Subsequent work is mainly concerned with two directions: better embeddings or optimization of interactive function.

Rather than using only IDs of users and items, some works \cite{koren2008factorization, chen2017attentive, he2018nais,wang2019kgat,ying2018graph, he2020lightgcn} focus on extending the representations of items or users by incorporating extensive ancillary information and historical user behaviors. Specifically, aSDEA\cite{dong2017hybrid} adds item attributes to assist learning of item representations while ACF\cite{chen2017attentive} and NAIS \cite{he2018nais} summarize user embeddings of historical items and treat them as user features. Furthermore, considering the effectiveness of constructing interactive data as a bipartite graph and the success of neural graph networks, more recent work such as NGCF \cite{wang2019neural}, PinSage\cite{ying2018graph}, and LightGCN\cite{he2020lightgcn},
reorganize personal histories in a graph and distill useful information from multi-hop neighbors to refine the embeddings.

On the other side, deep collaborative filtering models emphasize the importance of how user-item interactions are modeled. The inner product, widely used in MF algorithms, is replaced by a nonlinear function in neural networks\cite{rendle2012bpr, he2017neural, hsieh2017collaborative}. LRML \cite{tay2018latent} also attempts to use Euclidean distance as a metric to determine if there are interactions between users and items. 
\subsection{GCN for Recommendation}
Over recent years, graph convolutional networks have achieved great success in various fields, such as social network analysis \cite{qiu2018deepinf}, biomedical networks \cite{rhee2017hybrid, ma2019affinitynet, xu2019mr, wang2019mcne, wu2020graph} and recommender systems \cite{ying2018graph, wang2019neural, he2020lightgcn, wang2020disentangled}, for dealing with non-Euclidean data. Early works define graph convolution in the spectral domain based on graph Fourier transformation \cite{bruna2013spectral, defferrard2016convolutional}. Recent works, including GCN \cite{kipf2016semi}, GraphSAGE \cite{hamilton2017inductive} and GAT \cite{velivckovic2017graph}, redefine graph convolution in the spatial domain following the neighborhood aggregation scheme, which have shown superior performance in a wide range of tasks, such as node classification \cite{kipf2016semi}, link prediction\cite{li2020type}, graph representation learning \cite{ying2018hierarchical}, etc.   

In this paper, we concentrate on GCN models \cite{ying2018graph, wang2019neural, he2020lightgcn, chen2020revisiting, wang2020disentangled, wu2021self} in CF-based recommendation scenarios, where user-item interactive behaviors are formulated as a bipartite graph. Influential works involve PinSAGE \cite{ying2018graph}, which exploits efficient random walks for graph convolution to reduce the computational complexity in web-scale recommender systems, and NGCF \cite{wang2019neural}, which encodes high-level collaborative signals explicitly through propagating user and item embeddings on the graph. 
Later, Wang et al. \cite{wang2020disentangled} develop DGCF to model diverse user-item interactions and thus yield intent-aware representations. Recently, researchers have dedicated themselves to simplifying the design of GCN for recommendation tasks \cite{he2020lightgcn, chen2020revisiting, mao2021ultragcn}. Inspired by SGCN \cite{wu2019simplifying}, He et al. \cite{he2020lightgcn} propose LightGCN which designs a light graph convolution by removing both feature transformation and nonlinear activation for training effectiveness and generalization ability. For further burden reduction, UltraGCN \cite{mao2021ultragcn} devises a constraint loss to approximate infinite layers of message passing. Latterly, motivated by the power of contrastive learning, Wu et al. \cite{wu2021self} suggest a new learning scheme SGL, which takes node self-discrimination as a self-supervised task so as to moderate degree bias and enhance robustness to noisy interactions. 

Despite all these efforts, existing works learn user and item representations from their own neighborhood independently when predicting a user's preference for an item. That is, the user is unaware of the target item in the embedding process and vice versa. 
Our work addresses this problem and specifically integrates interactive guidance into graph convolution for the recommendation, which leads to interpretable interactive features and achieves significant performance gain.

\subsection{Attention Mechanism}
\label{sec:atten}
The attention mechanism, which is an attempt to focus on relevant things while ignoring others,  has been widely used in deep learning. The first attention model is proposed by Bahdanau et al. for machine translation \cite{bahdanau2014neural}, based on a simple but elegant idea that not only all input words but also their relative importance should be taken into account. Subsequent successful application fields include image captioning \cite{xu2015show}, entailment reasoning \cite{rocktaschel2015reasoning}, speech recognition \cite{chorowski2015attention}, and many more. 

% explain intra-attention and inter-attention which are more about learning task-independent representations.
Self-attention, first introduced in \cite{cheng2016long}, has been used across a range of tasks including reading comprehension, textual entailment, and abstractive summarization\cite{lin2017structured,parikh2016decomposable,paulus2017deep}, which mostly aims to finetune their own representations given context. In addition, attention could also depend on external information, which means it can board the limitation of using information from itself. Such an attention mechanism, termed inter-attention, usually involves task-specific information and then enables to focus on the parts most relevant to the final tasks \cite{luong2015effective,raffel2019exploring,luong2015effective,raffel2019exploring}.

The learning capacity of GCNs can also be improved by the attention mechanism. For instance, when performing neighborhood aggregation to refine a node's embedding, GAT \cite{velivckovic2017graph} assigns different importance to its neighbors following a self-attention strategy. Follow-up works, like GIN \cite{li2019graph} and KGAT \cite{wang2019kgat}, exploit the idea of GAT and successfully apply it to the recommendation field. While previous methods compute edge attention weights between two connected nodes, MAGNA \cite{wang2020multi} captures long-range node interactions by incorporating multi-hop neighboring context into attention calculation. 

Nevertheless, current graph attention methods are generally applied to graph structures with information-rich nodes, not suitable for user-item bipartite graphs where each node has ID features only. Moreover, the computation of attention coefficients is limited solely to the correlations between a central node and its neighbors, i.e., a self-attention mechanism. Hence, we design a novel and concise attention-based GCN architecture for the recommendation, which incorporates external target information to compute more purposeful and appropriate attention coefficients. 
% explain GAT, KGAT (application), GIN (application), multi-hop GAT,
% some variants like NGCF (application)
\section{InterActive Graph Convolutional Neural Network}
In this section, we first review the background and problem settings of recommendation in Section \ref{pre_GCN}. Then we describe the architecture of conventional GCN-based recommendation algorithms in Section \ref{GCN}.   
To address the weakness of existing GCN algorithms,
we propose the Interactive GCN in Section \ref{IA-GCN}.
Finally, a detailed comparison of IA-GCN and existing models are provided in Section \ref{model_analysis}. 
The notations are summarized in Table \ref{notations1}.

\subsection{Preliminaries}\label{pre_GCN}
Consider a typical recommendation system,
with $\mathcal U = \{u_1,...,u_n\}$ and  $\mathcal I =\{i_1,...,i_m\}$ as the sets of users and items respectively. The goal is to learn a function
$f:\mathcal U \times \mathcal I \to \mathbb R$ that predicts the preference score $\hat y_{u,i}$ of a \textit{target} user-item pair $(u,i)$.
Intuitively, an accurate predictor $f$ should assign a higher score to a \textit{positive} user-item pair $(u, i_+)$ with positive interactions (click, purchase, etc.), than to a \textit{negative} pair $(u,i_-)$ without positive interactions.

Following the Bayesian Personalized Ranking (BPR) \cite{rendle2012bpr}, the objective function is defined as,
\begin{equation}
\ell(\mathcal D) = -\frac{1}{|\mathcal D|}\sum_{(u,i_+,i_-)\in \mathcal D} \ln \sigma (\hat y_{u,{i_+}}-\hat y_{u,i_-})+\lambda R,
\end{equation}
where $\mathcal D$ is a dataset of triplets, triplet $(u,i_+,i_-)$ indicates that user $u$ prefers item $i_+$ to item $i_-$, $\lambda R$ is the regularization term on all model parameters and $\sigma$ is the sigmoid function.

In the widely used collaborative filtering (CF) setting, 
the preference score $\hat y_{u,i}$ is predicted as the inner product of embeddings of the target user $u$ and the target item $i$, namely,
\begin{equation}\label{eq:score0}
\hat y_{u,i} =f(u,i)=<\mathbf e_u^0, \mathbf e_i^0>,
\end{equation}
where $\mathbf e_u^0, \mathbf e_i^0 \in\mathbb R^d$ are the embeddings of $u$ and $i$.
Totally, we have $m+n$ embedding vectors, which would be randomly initialized and trained end-to-end together with other model parameters.

Although simple and efficient, CF is usually insufficient in capturing satisfactory embeddings for users and items. 
The key reasons are that CF only makes use of \textit{first-order connectivity} from user-item interactions, no \textit{high-order connectivity}, and this connectivity is modeled \textit{explicitly} only in the objective function \cite{wang2019neural}.

\subsection{Graph Convolution Framework} \label{GCN}
\begin{table}
\small
\caption{Important Notations Used in Section 3}\label{notations1}
\begin{center}
\begin{tabular}{ll|ll}
   \toprule
$\mathcal U,\mathcal I$ & user,item set & $n,m$ & \# users, \# items \\
$\hat y_{u,i}$ & predicted score & $f$ & prediction function\\
$\mathbb R$ & real number& $i_+,i_-$ & positive, negative item \\
$\mathcal D$ & triplet dataset  &$u,i$ & target user, item \\ 
$\sigma$ & sigmoid & $\lambda R$ & regularization\\
$\ell$ & loss function & $\mathbf e^0$ & original embedding\\
$K$ &total orders &$d$ & feature dimension \\
 $\mathcal G$ & graph & $\mathbf e^k$ & $k$-order feature  \\
$\mathcal V, \mathcal E$ &  vertex, edge &$\mathbf e^*$ & all order features   \\
$p$ & parent node & $\mathcal N$ & children /neighbor set \\
$c$ & child node & $Agg$ & Aggregator function \\
$g$ & guide node & $|g, |u, |i$ & conditional on $g,u,i$\\
 $<\cdot, \cdot>$ & inner product& $\alpha_{p,c}$ & $c$'s importance to $p$ \\ 
$Comb$ & Combination function& $\beta_{k}$ & importance of $k$-th layer\\ 
  & & $\tau$ & temperature parameter \\
 \bottomrule
\end{tabular}\end{center}
\end{table}

To address the limitation of CF, GCN-based recommendation algorithms \cite{wang2019neural, wang2019kgat} explicitly exploit the high-order connectivity among users and items through graph convolution.
Usually, the user-item interactions are formulated as an undirected bipartite graph $\mathcal G =(\mathcal V, \mathcal E)$, where
both users and items act as  graph nodes, i.e.,
 $\mathcal V = \mathcal U \cup  \mathcal I$ and  the $u-i$ interactions as edges, i.e., $\mathcal E  \subseteq \mathcal U \times \mathcal I$.
Then with the help of $\mathcal G$, the high-order connectivity is explicitly modeled. Specifically, the preference score is predicted from not only 1). $u$, $i$'s own embeddings $\mathbf e_u^0$, $\mathbf e_i^0$ that were used in traditional CF (Eq.\eqref{eq:score0}), but also 2). their  \textit{high-order} features, denoted as $\mathbf e_u^k, \mathbf e_i^k$, $k \in \{1,...,K\}$, where a $k$-order feature $\mathbf e_u^k$/$\mathbf e_i^k$ summarizes the information within $u$/$i$ 's $k$-hop neighborhood on graph $\mathcal G$. And the prediction score is calculated as,
\begin{equation}\label{eq:scorek}
\hat y_{u,i} =f(u,i)=<\mathbf e_u^*, \mathbf e_i^*>,
\end{equation}
where $e_u^*$ and $e_i^*$ represent high-order features after $K$ convolutional layers.
Inspired by ResNet\cite{he2016deep}, many researches \cite{xu2018representation,chen2020simple} have shown that using skip connection to combine GCN layers can efficiently address the oversmoothing issue. In a common paradigm, a combination operator is used to gather information from historical representations
$[e_u^0, e_u^1,\cdots, e_u^K]$, $[e_i^0, e_i^1,\cdots,e_i^K]$ and then Eq.\eqref{eq:scorek} can be expanded as:
\begin{equation}\label{eq:comb}
\mathbf e_u^* =Comb(\mathbf e_u^0,\mathbf e_u^1,...,\mathbf e_u^K) ,~~\mathbf e_i^* =Comb(\mathbf e_i^0,\mathbf e_i^1,...,\mathbf e_i^K) \in \mathbb R^{(K+1)d},
\end{equation}
where $Comb$ are arbitrary combination of $u$ and $i$'s 0 to $K$ order features. 

In literature, the high-order features of $u$/$i$
 are usually calculated by two tree-like structures
that are rooted at $u$/$i$ and consist of $K$ stacked graph convolutional layers, as shown  in Fig.\ref{arch}.  
Specifically, for any parent node $p$ in the two trees, its children set $\mathcal N_p$ are (sampled from) $p$'s direct neighbors in $\mathcal G$, and the $k+1$-order features of the parent $p$ are aggregated from children's $k$-order features in a graph convolutional operation,
\begin{equation}\label{eq:agg_general} 
\mathbf e_p^{k+1} ={Agg}( \mathbf e_c^k:c\in \mathcal N_p ),~~k \in \{0,...,K-1\}
\end{equation}
where $Agg$ is an aggregator function that combines the children' features.
This convolutional operation is used iteratively 
along the tree from bottom to top, resulting in $\mathbf e^*_u$ and $\mathbf e^*_i$ for the final preference prediction.

Although commonly used in the recommendation, this two-tree structure is originally inherited from GCN for classification on every single node.
That is why there is \textit{no} explicitly interactions between the two trees. 
Each tree extracts the \textit{general} portraits of $u$/$i$ independently until the final fusion, i.e., the CF layer.
This structure, however, is actually not suitable for a recommendation since the recommendation is substantially different from node classification: it is not $u$/$i$'s general portraits, but their interactive features, e.g., $u$'s consideration when choosing $i$ and $i$'s partial characteristic that attracts $u$, that really contribute to the preference prediction. 
Suffering from the suboptimal late fusion architecture, conventional GCN-based recommendation algorithms usually fail in modeling the interaction between the target user and target item and are thus ineffective in preference prediction.

\subsection{Interactive GCN for Recommendation} \label{IA-GCN}
We tackle the late fusion issue in conventional GCN-based recommendations by proposing a novel architecture, Interactive GCN (IA-GCN), that is
specially designed to model the user-item interactions.
\if 0
By building explicit interactions between the two trees, IA-GCN is able to 
 extract interactive high-order features that are expressive and effective in preference prediction.
 \fi
We start by explaining our key idea, ``interactive''. 
 
\subsubsection{What is ``Interactive''?} 
In conventional models (Eq.\ref{eq:scorek}),
the high-order features of each user/item are extracted independently of its corresponding item/user in the target $u-i$ pair. Thus they are always \textit{fixed} and \textit{universal}.
In other words, for any two different items $i\neq j$, the $\mathbf e^*_u$ used in predicting $\hat y_{u,i}$ and $\hat y_{u,j}$ is identical. And for any users $u \neq v$, $\mathbf e^*_i$ for predicting $\hat y_{u,i}$ and $\hat y_{v,i}$ is also identical.

%We now illustrate the drawbacks of these fixed and universal high-order features through a toy example. 
% \begin{figure*}
% \centering
%     \begin{minipage}[b]{0.33\textwidth}
%     \centering
%     \includegraphics[width=\textwidth]{books1.pdf}
%     \subcaption{}
%     \label{book1}
%     \end{minipage}
% \hfill\vline\hfill
% \begin{minipage}[b]{0.3\textwidth}
% \centering
% \includegraphics[width=\textwidth]{books2.pdf}
%     \subcaption{}
%     \label{book2}
% \end{minipage}
% \hfill\vline\hfill
% \begin{minipage}[b]{0.3\textwidth}
% \centering
% \includegraphics[width=\textwidth]{books3.pdf}
%     \subcaption{}
%     \label{book3}
% \end{minipage}
% \begin{minipage}[b]{0.4\textwidth}
% \centering
% \includegraphics[width=\textwidth]{book_label2.pdf}
%     \label{book_label}
% \end{minipage}
%   \caption{Illustration of the proposed IA-GCN. The arrowed lines represent the flow of information, while the dashed lines represent the guide. In \ref{book1} and \ref{book2}, the user tree is aggregated under the guidance of historical and speculative novels, respectively, and it can be clearly seen that the assigned weights of user interacted items vary greatly. When predicting a novel with a historical genre, higher attention is paid to the viewed historical novels. Similarly, the viewed speculative novels contribute more to infer the probability of interacting with a speculative novel. However, when it comes to prose, no item is given special attention.}
%   \label{fig:books}
% \end{figure*}
Consider a toy example where the target user $u$ who has purchased a smart phone and a skirt, denoted as $i_1$ and $i_2$ respectively. So $u$ has two children, and its feature is aggregated as
$\mathbf e_u^{k+1} = Agg(\mathbf e_{i_1}^k,\mathbf e_{i_2}^k)$. 
The question is how to design the aggregator, or more specifically to allocate the relative importance of $\mathbf e_{i_1}^k$ and $\mathbf e_{i_2}^k$, so that $\mathbf e^*_u$ is highly expressive of $u$'s interest on $i$.
Naturally, similar items share more latent factors that attract the user, so the child resembling the target $i$ would contribute more to the preference prediction of $i$.
If $i$ is a laptop which is similar to the smart phone $i_1$,
$\mathbf e_{i_1}$ deserves higher importance.
And if $i$ is a dress, we should assign higher importance to $\mathbf e_{i_2}$.
Unfortunately, the aggregators in conventional GCNs lack essential guidance from the other tree and are thus not able to preserve the precious target-specific information through each convolutional operation.
And this drawback not only degrades the aggregator in the toy example (user tree with level $1$) but also consistently affects all other aggregators in both trees.
%the aggregators  in conventional GCN-based recommendation mostly lack essential guidance from the other tree and thus discard the precious target specific information.  With only general information, these fixed and universal high-order features are not able to meet the above adaptability requirement.
%the fixed high-order features are not able to meet this adaptability requirement. 

To address this limitation, IA-GCN builds explicit interactions between the two trees:  
\begin{itemize}
    \item The target user $u$ guides the aggregations in the item tree, i.e., to emphasize children similar to $u$.
    \item The target item $i$ guides the aggregations in the user tree, i.e., to allocate high importance to children similar to $i$.
\end{itemize}
This interactive guidance enables IA-GCN to focus on target-specific information through each graph convolution. 
So the resulting high-order features of the target $u$/$i$ are not fixed, but conditional on its corresponding $i$/$u$ in the target user-item pair: $e_u^*|i$ and $e_i^*|u$.
% We reformulate our prediction as,
% \begin{equation}\label{eq:scorek_condition}
% \hat y_{u,i} =f(u,i)=<\mathbf e_u^*|i, \mathbf e_i^*|u>.
% \end{equation}

% A crucial remaining question is that how to measure the similarity between a child to aggregate and its guide, i.e., the root of the other tree. Since the two can be either \textit{homogeneous} (user-user, item-item) or \textit{heterogeneous} (user-item) in a bipartite graph, the similarity measurement is a nontrivial problem.
A crucial question is how to measure the similarity between a child to aggregate and its guide, i.e., the root of the other tree. 

%Note that a similar but much simplified idea has been adopted in the user behavior modeling field \cite{}, which predicts the preference from only a 1-level user tree and no item tree. While extracting the interactive features for GCN-based recommendation remains a non-trivial problem due to the complicated topology in our bipartite graph.

\subsubsection{Interactive Guide}\label{sec:ia_guide}
Consider a graph convolutional operation that calculates a parent node $p$'s high-order features by aggregating the features of its children $\forall c \in \mathcal N_p$, under the guidance of node $g$. Using the interactive guidance strategy, there are the two following cases:
\begin{itemize}
    \item $g=i$, and $c\in\mathcal N_u$, namely the target item guides the neighborhood aggregation in the user tree.
    shown in Fig.\ref{arch_inter}, \textcircled{1}.
    \item $g=u$, and $c\in\mathcal N_i$, namely the target user guides the neighborhood aggregation in the item tree.
    shown in Fig.\ref{arch_inter}, \textcircled{2}.
\end{itemize}

Specifically,
$c$'s importance when aggregated to $p$ is allocated according to $c$'s similarity/relatedness with the guide $g$. 
We list several considerations for this strategy.
First, $c$ and $g$'s high-order features, although available, may be noisy due to the neighborhood propagation. So we propose to
calculate the importance score only from
$c$ and $g$'s $0$-order features. Second, a simple inner product of $c$ and $g$'s embedding vectors should be feasible for computing attention coefficients. When $c$ and $g$ are homogeneous nodes, it serves as similarity measurement. When they are heterogenous, it quantifies the relatedness between user-item pairs.  Third, more children, i.e., larger $|\mathcal N_p|$, does not necessarily indicate that $p$ is more important. So the scale of $p$'s aggregated high-order features should not increase with $|\mathcal N_p|$. We thus would control the total scale of all similarities over $\forall c \in \mathcal N_p$.

Finally, we formulate a child
$c$'s importance to $p$ guided by $g$ as,

\begin{equation}\label{eq:ho}
% \begin{aligned}
\alpha_{p,c}|g = \frac{\exp (<\mathbf{e^0_g}, \mathbf{e_c}>/\tau)}{\sum_{c'\in\mathcal N_p}  \exp (<\mathbf{e^0_g}, \mathbf{e_{c'}}>/\tau)} 
% &= \frac{\exp \mathbf{(e_{gc}}/\tau)}{\sum_{c'\in\mathcal N_p} \exp (\mathbf{e_{gc'}}/\tau)} \\
% \end{aligned}
\end{equation}

% \begin{equation}
% \begin{aligned}
% e_{gc} = <\mathbf{e^0_g}, \mathbf{e^0_c}> - \max\limits_{c''\in\mathcal N_p } ( <\mathbf{ e^0_g}, \mathbf{e^0_{c''}}>)
% \end{aligned}
% \end{equation}

where $\tau$ is a temperature parameter.The softmax layer in Eq.\eqref{eq:ho} ensures $\sum_{c\in \mathcal N_p} \alpha_{p,c}|g =1$, which consists with the third consideration.

Note that this aggregation importance in IA-GCN is fundamentally different from existing attention mechanisms in GCN, e.g., GAT \cite{velivckovic2017graph}.
Our $\alpha_{p,c}|g$ depends on $c$'s similarity to $g$,
i.e., the interactive guidance from the other tree. 
While in existing attentions, $\alpha_{p,c}$ depends on $c$'s similarity to $p$. Since the knowledge used is still limited to $c$'s own single tree,  existing algorithms are not able to extract interactive features.

\subsubsection{Interactive Convolution}
\label{ia_convolution}
In previous sections, we focus on how to allocate the importance in the aggregator, now we dig into the design of the aggregator in Eq.\eqref{eq:agg_general}.

In literature, earlier GCN works mostly fall in the heavy pipeline: linear transformation, weighted sum pooling, and nonlinear activation \cite{he2017neural}. While recent research \cite{wu2019simplifying, chen2020revisiting} highlights the fact that a \textit{light} aggregator, e.g., weighted sum pooling, usually achieves state-of-the-art performance. 
We take the LightGCN for example. When aggregating $c\in \mathcal N_p$ to $p$, they use,
\begin{equation} 
\label{agg_lightGCN}
\mathbf e_p^{k+1} =\sum_{c\in \mathcal N_p}  \frac{1}{\sqrt{|\mathcal N_p||\mathcal N_c|}} \mathbf e_c^k,
\end{equation}
where their aggregation weight is a simple normalization based on information in $c$ and $p$'s own tree.

Since the focus of IA-GCN is to introduce interactive guidance between the two trees, we propose to follow the simple and proven effective weighted sum pooling aggregator. Using the proposed interactive guidance, our convolutional operation is formulated as,
\begin{equation} 
\label{eq: agg_ia}
\mathbf e_p^{k+1}|g =\sum_{c\in \mathcal N_p}  \alpha_{p,c}|g ~\mathbf e_c^k,
\end{equation}
where $\alpha_{p,c}|g$ is the interactive weight, defined in Eq.\eqref{eq:ho}
% and in Eq \ref{eq:he} for $c$ in level $2,4,...$.

Note that IA-GCN is an easy-plug-in module that theoretically could be applied to any GCN-based recommendation method. 
By multiplying our interactive weights, many existing algorithms will benefit from the learned user-item interaction knowledge.

\subsubsection{Layer Combination and Model Prediction}\label{sec:layer_combination}
Having introduced the way for message passing, we aggregate $k+1$-order features starting from the original embeddings $e^0$ by $Agg$ operator defined in Eq.\eqref{eq: agg_ia}. Then a combination operator mentioned in Eq.\eqref{eq:comb} is applied to gather influential information from sequential layers. Such $Comb$ operation can be reformulated as follows:
% \begin{equation} 
% \begin{aligned}
% \mathbf e_u^*|i &= Comb(e_u^0|i, e_u^1|i,\cdots, e_u^K|i) \\
% \mathbf e_i^*|u &= Comb(e_i^0|u, e_i^1|u,\cdots,e_i^K|u)
% \end{aligned}
% \end{equation}
% Specifically, $Comb$ in our work can be summarized as:
% \begin{equation} 
% \begin{aligned}
% \mathbf e_u^*|i &= \sum_{k=0}^{K}\beta_k e_u^k|i,\space \mathbf e_i^*|u = \sum_{k=0}^{K}\beta_k e_i^k|u
% \end{aligned}
% \label{eq:layer_comb}
% \end{equation}
\begin{equation} 
\mathbf e_p^*|g = Comb(e_p^0|g, e_p^1|g,\cdots, e_p^K|g) 
\end{equation}
Specifically, $Comb$ in our work can be summarized as:
\begin{equation}
\begin{split}
&\mathbf{e_p^*|g} = \sum_{k=0}^{K}\beta_k e_p^k|g \\
&\text{s.t.} \; \beta_k\geq 0 \;\& \sum_{k=0}^{K}\beta_k = 1 
\end{split}
\end{equation} 
% \label{eq:layer_comb}
% \mathbf &e_p^*|g = \sum_{k=0}^{K}\beta_k e_p^k|g \\
% &\text{s.t.} \; \beta_k\geq 0 \;\& \sum_{k=0}^{K}\beta_k = 1 \nonumber
% \end{align}
$\beta_k$ denotes the ratio/importance to gather information from $k$-order features. $\beta_k$ can be not only hyper-parameters tuned based on experts knowledge but also variables jointly learning with graph convolution layers. Like Eq.\eqref{eq:score0}, our prediction in consideration of the interactive guidance is as follows:
\begin{equation}\label{eq:scorek_condition}
\hat y_{u,i} =f(u,i)=<\mathbf e_u^*|i, \mathbf e_i^*|u>,
\end{equation}
which models the interaction between the target user and target item from an early stage, preserves the precious target-specific information through each convolutional operation, and makes a prediction of the interaction probabilities at the end.

% \subsection{Why Interactive, An Analysis}\label{Strength}
% Experimentally

% comparative experiments, relation to GAT

% attention is not always good, bias may harm model performance, an interactive guide is the key.

% Theorectically (Need improvement)

% Interpretability of our algorithm

% external attention has shown effectiveness in NLP tasks

% GIN, why not take one step further

% KGAT

\subsection{Model Analysis}
\label{model_analysis}
In this subsection, we will discuss the similarities and differences between IA-GCN and existing models and provide deeper insights into the rationality of our design.

\subsubsection{\textbf{Relation with LightGCN}}
\label{relation_lightgcn}
For both LightGCN \cite{he2020lightgcn} and IA-GCN, the whole trainable parameters are 0-order embedding vectors $\{\mathbf e_u^0, \mathbf e_i^0 | u \in \mathcal U, i \in \mathcal I \}$ and layer combination coefficients $\{\beta_0, ...,  \beta_K\}$. That is, the model size of IA-GCN is exactly identical to LightGCN. Regarding model design, the only main difference is the way to aggregate the features of neighboring nodes ($cf.$ Section \ref{ia_convolution}). LightGCN uses the static degrees of parent $p$ and child $c$ to normalize the embeddings ($cf.$ Eq.\eqref{agg_lightGCN}), while IA-GCN computes dynamic attention scores according to child $c$'s similarity/relatedness with guide $g$ from the other tree ($cf.$ Eq.\eqref{eq: agg_ia}). Under fair experimental settings, IA-GCN consistently outperform LighGCN in terms of recommendation accuracy (evidence from table \ref{tb:overall_results}) and convergence rate (evidence from Fig.\ref{fig: converge}). Moreover, our model also has better interpretability since target information is explicitly encoded in attention coefficients. 

% \subsubsection{Relation with Link Prediction}

\subsubsection{\textbf{Relation with GAT}}
Although the core idea of GAT \cite{velivckovic2017graph} is consistent with IA-GCN, e.g., to learn a weighted aggregation of node features in a graph convolution, its implementation is fundamentally different from ours. GAT has feature transformation and non-linear activation operations, making it yield bad performance for CF-based recommendation. Thus, we will not present its results in the experimental section. Moreover, GAT follows a self-attention mechanism, while IA-GCN follows a inter-attention mechanism ($cf.$ Section \ref{sec:ia_guide}). Furthermore, IA-GCN has stronger expressive power. If we have $m$ user and $n$ items in total and want to make predictions for their interactions, GAT only yields $m$ user representations and $n$ item representations independently. Nonetheless, IA-GCN takes each user-item combination into consideration, and generates $mn$ user representations and  $nm$ item representations, with limited model parameters ($cf.$ Section \ref{relation_lightgcn}). 

\subsubsection{\textbf{Relation with GIN}}
To predict click-through rate in sponsored search, GIN \cite{li2019graph} constructs a co-occurrence commodity tree for each commodity in the user's real-time behaviors. Initially, it exploits the idea of GAT \cite{velivckovic2017graph} to aggregate those trees and yield high-level commodity representations, then weights the commodities in the user's behavior sequence according to its similarity with the target one. Though GIN also uses an external guide for neighborhood aggregation, it distinguishes from our IA-GCN from the following two aspects: 1) In GIN, the target commodity only provides the guidance to first-order neighbors in the user graph, whereas, IA-GCN guides neighborhood aggregation from all layers; 2) The guidance in GIN is unilateral from the target item to the user side, while IA-GCN can build bilateral interactive guidance ($cf.$ Fig.\ref{fig:teaser}). 

%While for conciseness, we will  omit the condition and slightly abuse our Eq (\ref{eq:scorek}) when no ambiguity is caused.

\section{Experiments}
We first compare our proposed IA-GCN with LightGCN and the other various state-of-the-art GCN-based CF algorithms. 
Afterwards, we present experiments for ablation study to illustrate the impact of the layer combination and the importance of reasonable guidance to justify the rationality of the design choices of IA-GCN. 
\subsection{Experimental Settings}
\subsubsection{Dataset Description}
As shown in Table ~\ref{tb:dataset}, we use three publicly available datasets: Gowalla, Yelp2018, and Amazon-Book, released by NGCF\cite{wang2019neural}. To keep the comparison fair, we closely follow and use the same data split as LightGCN\cite{he2020lightgcn}. In the training phase, each observed user-item interaction is treated as a positive instance, while we use negative sampling to randomly sample an unobserved item and pair it with the user as a negative instance.
\begin{table}[ht]
\centering
\caption{Statistics of the datasets}
\label{tb:dataset}
\begin{tabular}{l|l|l|l|l} 
\hline
Dataset     & \#Users & \#Items & \#Interactions & Density  \\ 
\hline
\hline
Gowalla     & 29,858  & 40,981  & 1,027,370      & 0.00084  \\ 
\hline
Yelp2018    & 31,668  & 38,048  & 1,561,406      & 0.00130  \\ 
\hline
Amazon-Book & 52,643  & 91,599  & 2,984,108      & 0.00062  \\
\hline
\end{tabular}
\end{table}
\subsubsection{Compared Methods}\label{sec:com_method}
We compare IA-GCN with the state-of-the-art methods and investigate how the guidance affects the performance, so we group them in terms of attention mechanism covering three groups: \textbf{no attention} (GC-MC, DisenGCN, LightGCN, and DGCF), \textbf{intra-attention or its variants} (NGCF, NGCF$_{light}$ and SA-GCN).
\begin{itemize}
\item {\textbf{GC-MC}}\cite{berg2017graph}: This formulates matrix completion task as a link prediction task and proposes an auto-encoder framework on the bipartite user-item interaction graph, where the encoder obtains representations of nodes and the decoder reconstructs the rating links.
\item {\textbf{DisenGCN}}\cite{ma2019disentangled}: This exploits the neighbor routing mechanism which can dynamically disentangle the latent factors behind the graph edges and learn disentangled node representations.
\item {\textbf{LightGCN}}\cite{he2020lightgcn}: This simplifies GCNs for collaborative filtering by removing feature transformation and nonlinear activation keeping only neighborhood aggregation, which is stated as the most essential component.
\item {\textbf{DGCF}}\cite{wang2020disentangled}: This emphasizes the importance of differentiating user intents on different connected items and disentangles the latent user intents at a more granular level.
\item {\textbf{NGCF}}\cite{he2017neural}: This method adopts three GCN layers on the user-item interaction graph, aiming to refine user and item representations via the same propagation rule: feature transformation, neighborhood aggregation, and nonlinear activation. 
Moreover, it can be viewed as \textbf{a variant of self-attention mechanism}, since there exists a representation interaction term that makes the message being propagated dependent on the affinity between a child node and its parent.
% by incorporating dependent on the affinity between neighbors.
\item {\textbf{NGCF$_{light}$}}: This follows all the settings of NGCF especially inner products between connected nodes when propagating embeddings. However, it removes the feature transformation matrices and non-linear activation function to keep consistent with LightGCN.
\item {\textbf{SG-GCN}}: This follows all the settings of our proposed IA-GCN algorithm except the choice of guides $g$ mentioned in Eq.\eqref{eq:ho}. Rather than using the interactive guidance, SG-GCN(short for \underline{S}elf-\underline{G}uided GCN) aggregates trees guided by their own roots. 
% \item {\textbf{NIA-GCN}}\cite{sun2020neighbor}: This captures the relationship between individual neighbor-neighbor pairs at each GCN layer explicitly and, because of the heterogeneous nature of the user-item bipartite graph, independently aggregates various entities at different layers to obtain different information.
% \end{itemize}
% To investigate how the guidance affects the performance, we also consider the followings methods:
% \begin{itemize}
\end{itemize}
We further compare IA-GCN with several methods including no attention methods(LightGCN) and intra-attention methods or its variants (NGCF, NGCF$_{light}$ and SA-GCN) to validate our rational choice of guidance. The detailed analysis can be found in Section \ref{sec:gc}.

\subsubsection{Evaluation Metrics}
To evaluate top-N recommendation, Recall@20 and NDCG@20 are chosen for its popularity in GCN-based CF models\cite{he2017neural,he2020lightgcn}. When testing, we regard the items that the users in the test set interacted with as the positive ones and evaluate how the positive items rank among all other un-interacted items. The average results w.r.t. the metrics over all users are reported.

\subsubsection{HyperParameter Settings}
Same as LightGCN, the embedding size is fixed to 64 for all models and the embedding parameters are initialized with the Xavier \cite {glorot2010understanding} method. We optimize IA-GCN with the Adam \cite{kingma2014adam} optimizer and the default mini-batch size of 1024 (on Amazon-Book, we adapt a mini-batch size of 2048, which follows the setting of LightGCN). 
Learning rate is searched in the range of \{$5e^{-4}$, $5.5e^{-4}$, $6e^{-4}$, \ldots, $1e^{-4}$\} in view of our validation gap and convergence rate. We choose $1e^{-4}$ as the $L_2$ regularization coefficient $\lambda$ and the early stopping and validation strategies are the same as LightGCN. 
For fair comparison with other methods, we follow the same setting for layer combination and set $\beta_k$ (Eq.\eqref{eq:comb}) uniformly as $1/(K+1)$. 
% Moreover, we also conduct experiments to explain the impact of layer combination on our proposed IA-GCN, which is further discussed in Sec \ref{sec:lc}.
%The final representation of a user or an item obtains from the embeddings at each layer.
\subsection{Performance Comparison with LightGCN}
We compare LightGCN with our IA-GCN by exhaustively reporting the results at different layers in Table ~\ref{tb:layer_results}. The comparison with other algorithms shown in this table will be discussed in Section \ref{sec:gc}. Our discussion is as follows:
\begin{itemize}
    % \item  IA-GCN consistently yields better performance than LightGCN on three datasets when the layer number ranges from 1 to 3, indicating the superiority of our proposed model. We attribute such performance to the following: 1). attention mechanism denoise the messages passing through propagation operation, which purify the information and improve the final representation to some extent. The edge weights for the learning of user-item relationships in IA-GCN are also more reasonable; 2). interactive guidance advances the fusion procedure between user tree and item tree to avoids sub-optimal late fusion when aggregating independently; 3). explicit interactions between the two trees in recommender systems is essential and this interactive guidance enable GCNs to focus on target-specific information through every convolution and thus capture interactive characteristics in high-order features.
    % 整体来看效果
    \item IA-GCN consistently outperforms LightGCN on three benchmark datasets when the number of layers ranges from 1 to 3, demonstrating the effectiveness of our proposed method. We attribute such performance gain to the following reasons: 1) LightGCN is vulnerable to user-item interaction noises, while the attention mechanism in IA-GCN could help mitigate the negative impact brought by latent noisy interactions and improve the representation learning; 2) LightGCN suffers from the sub-optimal late fusion of user and item features, whereas, IA-GCN fully integrate user and item trees by building interactive links between the two ($cf.$ Fig. \ref{arch_inter}); 3) IA-GCN aggregates a user/item tree in different ways when facing different target items/users, while LightGCN performs neighborhood aggregation in a uniform manner and thus lose target-specific information. 
    % \item The highest Recall@20 reported by LightGCN on Amazon Book is 0.0411, whereas, our IA-GCN increases the metric by 15.2\%. Although the relative improvement on Gowalla and Yelp2018 is not as dramatic as that on Amazon-book, IA-GCN shows its superiority and effectiveness starting from a lower layer number. Specifically,  3-layer's LightGCN underperforms 1-layer IA-GCN on Amazon-Book and Yelp2018 and 2-layer's IA-GCN on Gowalla according to Recall@20.
    % 从左往右来分析结果（layer by layer）
    \item Increasing the model depth from 1 to 3 is able to improve the performance in most cases yet it reaches a plateau afterward. This observation is consistent with LightGCN's finding.
    % 从上到下来分析结果（group 之间进行对比）
    \item The highest Recall@20 reported by LightGCN on Amazon Book is 0.0411, whereas, our IA-GCN increases the metric by 15.2\%. The relative improvement on Gowalla and Yelp2018 is not as dramatic as that on Amazon-book, which might be caused by the natural metrics of the datasets: the density as shown in Table \ref{tb:dataset}. 
    CF algorithms suffer from the data sparsity problem\cite{grvcar2005data}, that is, the users' preference data on items are usually too few and too unreliable to reflect the users’ true preferences. 
    Supervision signals from the other tree help IA-GCN to perform neighborhood propagation with preferences, in which the sparsity problem can be alleviated to some extent.
\end{itemize}
\begin{table*}
\small
\centering
% NGCF, NGCF$_{light}$ and SG-GCN aggregate neighbors guided by its own node while IA-GCN involves the corresponding tree when aggregating. 
\caption{Performance comparison between IA-GCN and competing methods at different layers.
Algorithms are divided into three groups: No-Att, Intra-Att, and Inter-Att. The relative improvement reported is compared with Light-GCN.}
\label{tb:layer_results}
\begin{minipage}[]{\linewidth}
\centering
\subcaption{Gowalla}
\begin{tabular}{@{}clllllll@{}}
\toprule
& Num Layers 
& \multicolumn{2}{c}{1 Layer} 
& \multicolumn{2}{c}{2 Layers}
& \multicolumn{2}{c}{3 Layers} \\ \midrule
& Method  & recall & ndcg   & recall & ndcg   & recall & ndcg  \\ \midrule
\multirow{1}{*}{No-Att} 
& LightGCN & 0.1755 & 0.1492 & 0.1777 & 0.1524 & 0.1823 & 0.1555\\ \midrule
\multirow{3}{*}{Intra-Att} 
& NGCF  & 0.1556 & 0.1315 & 0.1547 & 0.1307 & 0.1569 & 0.1327 \\
& NGCF$_{light}$ & 0.1605 & 0.1300 & 0.1635&  0.1333 & 0.1621  & 0.1331  \\
& SG-GCN & 0.1673 & 0.1440 & 0.1745 & 0.1498 &0.1741 &  0.1498 \\ \midrule
Inter-Att 
& IA-GCN & \textbf{0.1762}(+0.40\%) & \textbf{0.1509}(+1.13\%)  & \textbf{0.1821}(+2.45\%) & \textbf{0.1551}(+1.77\%)  & \textbf{0.1839}(+0.90\%) & \textbf{0.1562}(+0.46\%) \\
\bottomrule
\end{tabular}
\label{tb:gowalla}
\end{minipage}
% \vspace{0.8em}
\begin{minipage}[]{\linewidth}
\centering
\subcaption{Yelp2018}
\begin{tabular}{@{}clllllll@{}}
\toprule
& Num Layers 
& \multicolumn{2}{c}{1 Layer} 
& \multicolumn{2}{c}{2 Layers}
& \multicolumn{2}{c}{3 Layers} \\ \midrule
& Method  & recall & ndcg   & recall & ndcg   & recall & ndcg  \\ \midrule
\multirow{1}{*}{No-Att}    
& LightGCN   & 0.0631       & 0.0515       & 0.0622        & 0.0504       & 0.0639        & 0.0525       \\ \midrule
\multirow{3}{*}{Intra-Att} 
& NGCF       & 0.0543       & 0.0442       & 0.0566        & 0.0465       & 0.0579        & 0.0477       \\
& NGCF$_{light}$ & 0.0596 & 0.0482 & 0.0578 &  0.0470 & 0.0562  & 0.0456  \\
& SG-GCN      & 0.0624 & 0.0516 & 0.0634  & 0.0522  &  0.0650  & 0.0535     \\ \midrule
Inter-Att                  
& IA-GCN     & \textbf{0.0647}(+2.47\%)       & \textbf{0.0532}(+3.28\%)       & \textbf{0.0657}(+5.56\%)        & \textbf{0.0535}(+6.22\%)       & \textbf{0.0659}(+3.13\%)           & \textbf{0.0537}(+2.22\%)          \\  
\bottomrule
\end{tabular}
\label{tb:yelp}
\end{minipage}
% \vspace{0.8em}
\begin{minipage}[]{\linewidth}
\centering
\subcaption{Amazon-Book}
\begin{tabular}{@{}clllllll@{}}
\toprule
& Num Layers 
& \multicolumn{2}{c}{1 Layer} 
& \multicolumn{2}{c}{2 Layers}
& \multicolumn{2}{c}{3 Layers} \\ \midrule
& Method  & recall & ndcg   & recall & ndcg   & recall & ndcg  \\ \midrule
\multirow{1}{*}{No-Att}    
& LightGCN   & 0.0384       & 0.0298       & 0.0411        & 0.0315       & 0.0410        & 0.0318       \\ \midrule
\multirow{3}{*}{Intra-Att} 
& NGCF       & 0.0313       & 0.0241       & 0.0330        & 0.0254       & 0.0337        & 0.0261       \\
& NGCF$_{light}$ &  0.0371 & 0.0281 & 0.0368 &   0.0279 & 0.0367 & 0.0278    \\
& SG-GCN      & 0.0401       & 0.0314       & 0.0404        & 0.0315       &   0.0405    &    0.0316    \\ \midrule
Inter-Att                  
& IA-GCN     & \textbf{0.0463}(+20.65\%)     & \textbf{0.0364}(+22.00\%)       & \textbf{0.0450}(+9.37\%)        & \textbf{0.0350}(+11.21\%)       & \textbf{0.0472}(+15.20\%)           & \textbf{0.0373}(+17.20\%)     \\  
\bottomrule
\end{tabular}
\label{tb:amazon}
\end{minipage}
% \begin{tablenotes}
% \footnotesize
% \begin{itemize}
% \item[*] The scores of competing methods are directly copied from the published papers.
% \end{itemize}
% \end{tablenotes}
\end{table*}
\begin{figure*}
    \centering
    \includegraphics[width=\textwidth]{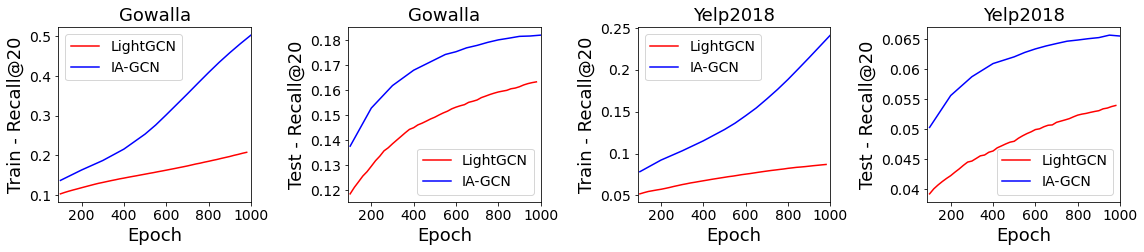}
    \caption{Training and Testing evaluation metrics curves of LightGCN and IA-GCN. All methods are evaluated by Recall@20 ranging from 100 to 1000 epochs on Yelp2018. For fairness comparison, we conducted experiments on Gowalla and Yelp2018 under the learning rate $5.5e^{-5}$ and $6.5e^{-5}$ respectively. }
    \label{fig: converge}
\end{figure*}
We further plot the training metrics and test metrics on the same graph(Fig.\ref{fig: converge}) to illustrate the effectiveness and efficiency of our IA-GCN. we omit the performance w.r.t. NDCG which has a similar trend. The observations and analysis are as follows:
\begin{itemize}
    % 训练的趋势
   \item During the entire training process shown in Fig.\ref{fig: converge}, IA-GCN unfailingly obtains high training evaluation metrics, indicating that our model classifies the training data better than LightGCN. Impressively, at the first 100 epochs, IA-GCN achieved 0.075 Recall@20 on training data while LightGCN reached the same evaluation metrics by epoch 1000. 
   % 测试的趋势
    \item A great generalization power enables IA-GCN to transfer from training superiority to better test performance. From the metrics evaluated on test data, IA-GCN surpasses LightGCN by a remarkable margin. Inferring from the trend, IA-GCN already converged within1000 epochs under such a small learning rate ($5.5e^{-5}$ or $6.5e^{-5}$) yet it still takes time for LightGCN to converge.
\end{itemize}
\subsection{Performance Comparison with State-of-the-Arts}
\begin{table}[]
\centering
\caption{Overall Performance Comparison}
\label{tb:overall_results}
\small
\begin{tabular}{l|cc|cc|cc}
% NIA-GCN& 0.1726& 0.1358 & 0.0599& 0.0491 & 0.0369& 0.0287         \\
\hline
  & \multicolumn{2}{c|}{Gowalla}  & \multicolumn{2}{c|}{Yelp2018} & \multicolumn{2}{c}{Amazon-book} \\ 
   & recall  & ndcg   & recall  & ndcg          & recall         & ndcg           \\ \hline
\hline
GC-MC& 0.1395& 0.1204 & 0.0462& 0.0379 & 0.0288& 0.0224         \\
DisenGCN & 0.1356& 0.1174 & 0.0558& 0.0454 & 0.0329& 0.0254     \\
NGCF& 0.1569& 0.1327 & 0.0579& 0.0477 & 0.0337& 0.0261         \\
LightGCN & 0.1823 & 0.1555 & 0.0639& 0.0
525& 0.0411  & 0.0318    \\ 
DGCF  & \textbf{0.1842} & \underline{0.1561} & \underline{0.0654} & \underline{0.0534} & \underline{0.0422} & \underline{0.0324} \\ 
% \midrule
% NGCF$_{light}$& 0.1635& 0.1333 & 0.0596& 0.0482 & 0.0371& 0.0281         \\
% SA-GCN & 0.1745& 0.1498 & 0.0650& 0.0535 & 0.0405& 0.0316     \\
% \midrule
IA-GCN & \underline{0.1839} &  \textbf{0.1562} &\textbf{0.0659}& \textbf{0.0537}      & \textbf{0.0472} &\textbf{0.0373}  \\ \hline
\hline
% \%improv. & -0.02\% & +0.06\%   &    +0.61\%    &  +0.37\%&        +11.85\%  &    +15.12\%  \\ \hline
% $p$-value. &                      &        &               &               &                &                \\ \hline
\end{tabular}
% \begin{tablenotes}
% \footnotesize
% \begin{itemize}
% \item[*] The scores of other algorithms are copied from published papers.
% \item[*] Bolded \textbf{items} are the best result at each case and the underlined \underline{item} represents the second best.
% \end{itemize}
% \end{tablenotes}
\end{table}
% Considered that IA-GCN already outperforms the state of the arts among 3 layers, we terminate our experiments at 4 layers.
 An overall comparison between IA-GCN with other competing methods is shown in Table ~\ref{tb:overall_results}, which reports the best performance obtained within \textbf{3} convolutional layers of each method. We omit the performance of NGCF$_{light}$ and SG-GCN here ($cf.$ \ref{sec:gc} for more analysis.
 Note that IA-GCN can be further improved by adding neglectable parameters to tune the importance of layers ($cf.$ Section \ref{sec:lc}), while here we only use vanilla layer combination ($\beta_k = 1/(K+1)$). In most cases, our IA-GCN achieves \textbf{significant improvements} over all other methods, demonstrating its rationality and effectiveness.

 % GC-MC不行因为它只用了一阶邻居（而我们用了3阶）
 % NGCF 比GC-MC好，因为它高阶；LightGCN好是因为light -> 我们延续了好的结构。
GC-MC performs poorly on three datasets, which might suggest that it is usually insufficient in capturing satisfactory embeddings for users and items when only utilizing the first-order connectivity. Compared with DisenGCN and NGCF, LightGCN substantially performs better on three datasets, which is consistent with their claim. The reason might be the statement in LightGCN paper: too heavy feature transformation and non-linear activation might be harmful to the final performance. Both explain the reasonable design of our IA-GCN since we employ \textbf{high-order neighbors} and \textbf{light propagation module} without the complicated non-linear feature transformations.
% Moreover, this conclusion can also explains the reason why our modified NGCF$_{light}$ outperforms NGCF in all cases.
% However, IA-GCN consistently surpasses LightGCN in all cases, which attributes to the interactive features extracted by inter-attention mechanism.

 % DGCF有效，有效原因是它提纯了信息，和我们attention一致的思路；
 DGCF serves as the strongest baseline in all cases and this verifies the high effectiveness of its disentangling module and propagation mechanism. Both of them aim to purify and distill helpful information from all high-order features, which is also the natural merit of the attention mechanism applied in IA-GCN. Although the performance of IA-GCN is on par with DGCF on Gowalla, IA-GCN surpasses it on the other two datasets especially 15\% relative improvement on Amazon Book. This phenomenon further verifies that the \textbf{attention mechanism} between two trees mitigates the data sparsity problem.
 
Since IA-GCN's improvements mainly attribute to the early fusion for the item and user trees and attention mechanism to degrade irrelevant items, its ideas are likely to incorporate with some state-of-the-art models. Theoretically, it could be applied to any GCN-based recommendation method as \textbf{an easy-plug-in module}. For instance, on top of DGCF, it is possible to consider the guidance from the other tree and disentangle the latent user intents.
\subsection{Study of IA-GCN}
\begin{figure*}[htbp!]
     \centering
     \begin{subfigure}[b]{0.32\textwidth}
         \centering
         \includegraphics[width=\textwidth]{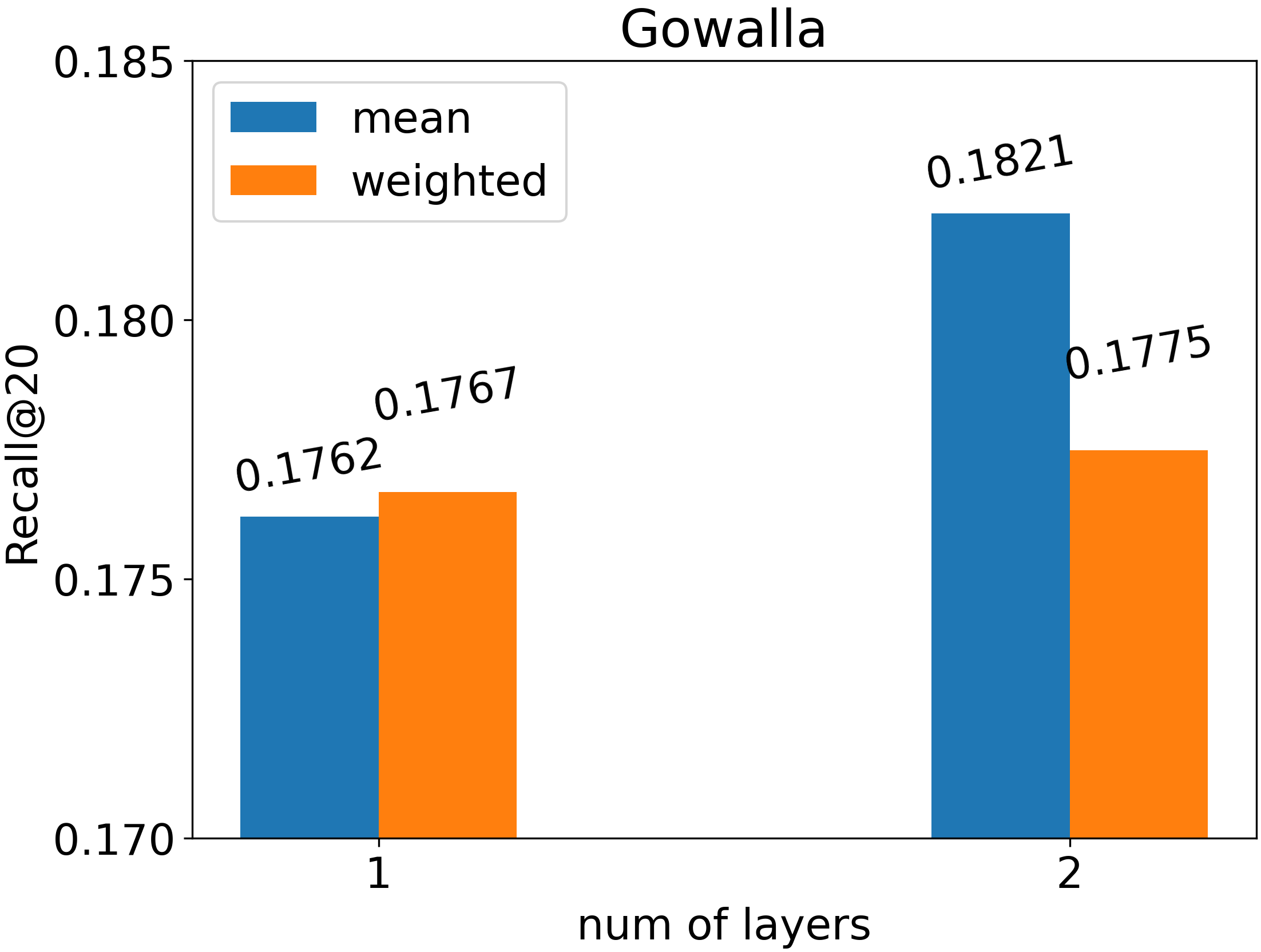}
         \label{fig: lc1}
     \end{subfigure} 
     \hfill
     \begin{subfigure}[b]{0.32\textwidth}
         \centering
         \includegraphics[width=\textwidth]{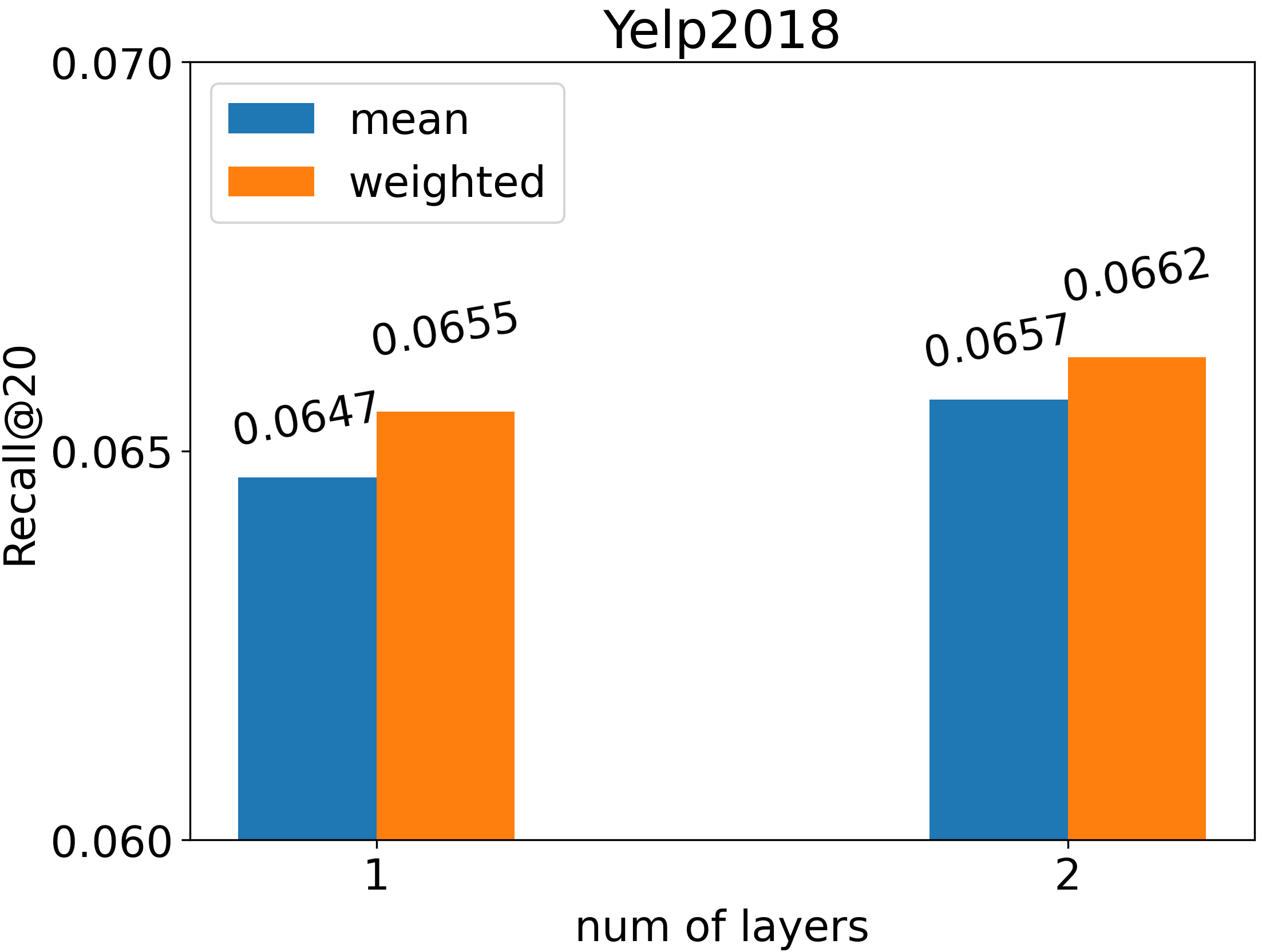}
         \label{fig: lc2}
     \end{subfigure}
     \hfill
     \begin{subfigure}[b]{0.32\textwidth}
         \centering
         \includegraphics[width=\textwidth]{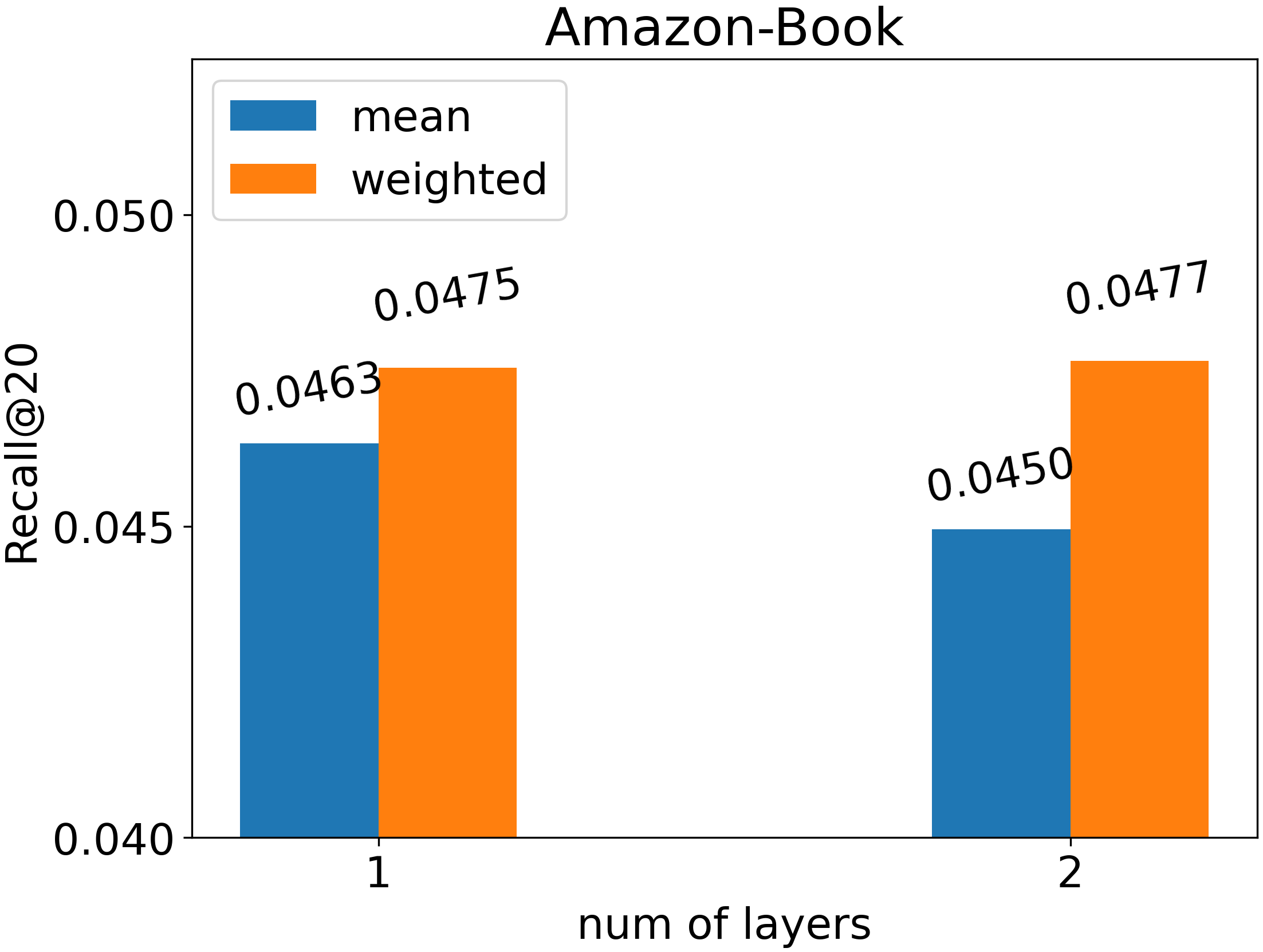}
         \label{fig: lc3}
     \end{subfigure}
    \caption{Results of IA-GCN and the variant that uses weighted layer combination at different layers on Gowalla, Yelp2018, and Amazon-Book}
    \label{fig: lc}
\end{figure*}
% We perform ablation studies to show the effects of layer combinations. For verifying the appropriate guidance mentioned in Section \ref{sec:layer_combination}, further experiments reveal the rationality of our method. 
In this section, we analyze the impact of layer combinations. To justify the choice of the interactive guide in Section \ref{sec:ia_guide}, we compare IA-GCN with several self-attention-based GCN algorithms. 
\subsubsection{Impact of Layer Combination}\label{sec:lc}
% \begin{figure*}[htbp!]
%      \centering
%      \begin{subfigure}[b]{0.32\textwidth}
%          \centering
%          \includegraphics[width=\textwidth]{lc1.png}
%          \label{fig: lc1}
%      \end{subfigure} 
%      \hfill
%      \begin{subfigure}[b]{0.32\textwidth}
%          \centering
%          \includegraphics[width=\textwidth]{lc2.png}
%          \label{fig: lc2}
%      \end{subfigure}
%      \hfill
%      \begin{subfigure}[b]{0.32\textwidth}
%          \centering
%          \includegraphics[width=\textwidth]{lc3.png}
%          \label{fig: lc3}
%      \end{subfigure}
%     \caption{Results of IA-GCN and the variant that uses weighted layer combination at different layers on Gowalla, Yelp2018 and Amazon-Book}
%     \label{fig: lc}
% \end{figure*}
We further conduct experiments to explore a better layer combination way. As described in Section ~\ref{sec:layer_combination}, ``$weighted$'' denotes a weighted sum of layer embeddings according to their importance $\beta_k$ learned during the model training,  while ``$mean$'' denotes the vanilla combination approach: $\beta_k= 1/(K+1)$. It is worthwhile to notice that the trainable parameters are only the original embeddings $\mathbf{e}^0$ in ``$mean$'' but $K+1$ extra layer combination parameters in ``$weighted$''.

Fig.\ref{fig: lc} gives a detailed comparison of three datasets to show the impact of layer combination coefficients. We find out that:
\begin{itemize}
    \item Apart from 2-layer Gowalla,
    % the gradual decrease when switches to ``$weighted$''  on 2-layer Gowalla,
    ``$weighted$'' brings an improvement in all other situations. This reveals the varying contributions of multi-order features to the final performance. 
    \item On the Amazon-Book dataset, we find that its performance drops quickly when increasing the number of layers from 1 to 2 when using ``$mean$'' layer combination, whereas, the 2-layer Recall@20 increases from 0.0450 to 0.0477 via re-weighting the features aggregated from different layers. This implies that reasonable layer weights could help distill information from high-order neighbors more effectively. 
    % This reveals that the decrease here is not over-smoothing issues caused by involving high-order neighbors. (there is) 
    % Appropriate combination and combination ratio $\beta$ can further improve the performance.
\end{itemize}
\subsubsection{Impact of Guide Choice}\label{sec:gc}

As shown in Table \ref{tb:layer_results} and Fig. \ref{fig: converge}, the inter-attention mechanism introduced in IA-GCN can improve model expressiveness and accelerate convergence. These benefits are brought by interactive guidance links between user-item pairs as presented in Fig. \ref{arch_inter}. Supplementary experiments using self-attention methods or their variants (described in Section \ref{sec:com_method}) further prove the necessity of introducing external guidance. We have the following observations:

\begin{itemize}
\item Clearly, encoding affinities between parent nodes and children nodes are harmful. NGCF and NGCF$_{light}$ model such affinities using the inner product and subsequently underperform LightGCN. It is noticed in Table \ref{tb:layer_results} that on Yelp2018 and Amazon-Book, the performance of NGCF$_{light}$ gets worse when increasing the number of layers. This implies that this affinity term negatively affects the representation learning and act as noise. Hence, when we increase the model depth and incorporate more neighbors, we actually introduce more noises and thus degrade model training. 

\item SG-GCN only has one difference from IA-GCN. It aggregates a user/item tree under the guidance of its own root instead of the target root. SG-GCN surpasses LightGCN on 2- or 3-layer Yelp2018 and 1-layer Amazon-Book but fails in the rest, while IA-GCN exhibits significant performance gain over LightGCN in all the experiments, as shown in Table \ref{tb:layer_results}. 
% In this way, we pay more attention  However, the obtained representations only represent the general portraits of users/items. 
\end{itemize}

% \begin{itemize}
% % 分析组间的关系，LightGCN> intra-attention, 但< inter-Att，说明interactive guidance才是关系，attention不一定都是适合的好的，合适的guidance才能够有好的效果
%     \item In most cases, LightGCN outperforms ``Intra-Att'' methods especially the NGCF-based ones while SG-GCN surpasses LightGCN on 2- or 3-layer Yelp2018 and 1-layer Amazon-Book and sometimes on par with it. Nonetheless, IA-GCN is the best in all cases. 
%     When aggregating one tree, LightGCN treats all nodes at each layer equally and iteratively passes the message from lower to higher yet all edges are given different importance in attention-based algorithms depending on the similarity between the node and the guidance. For ``Intra-Att'', they are likely to align some nodes from one tree, which hinders information representation on the nodes that are less related to the guide. The situation is different in ``Inter-Att'' since IA-GCN employs attention interacted with another tree. This leads to interactive and interpretable features, effectively distilling target-specific information through each graph convolutional operation. 
% \end{itemize}
% \subsubsection{Smoothness Analysis}
% % TODO http://proceedings.mlr.press/v119/wang20k/wang20k.pdf
% Describe over-smooth

% refer contrastive learning the alignment between positive pairs.
\section{CONCLUSION AND FUTURE WORK}
% \section{CONCLUSION}
% In this work, we propose a novel graph attention framework IA-GCN to address a long-overlooked issue, the sub-optimal late fusion of user and item features in Collaborative Filtering. When predicting a user's preference for an item, 
% the user tree, and the corresponding item tree will be aggregated from bottom to top under the guidance of each other, to extract interactive features as well as emphasize target-specific information. We conduct various experiments to demonstrate the superiority of our model, in terms of recommendation effectiveness, generalization ability, and training convergence. 

% Our work represents an initial attempt towards dynamic or interactive graph convolution for recommendation and opens up a new research line. In future work, we will follow the direction of the attention mechanism and exploit various attention mechanisms e.g. extending a single guide to muti-guides. Further, we will incorporate more side information like item attributes\cite{dong2017hybrid} to assist measurement of attention. 
In this work, we point out a long-overlooked issue in Collaborative
Filtering: the representations of the target user and the target item
are generated independently which results in sub-optimal preference prediction. To address this issue, we propose to introduce interactions between the user tree and the item tree by emphasizing the target information of each other. By this means, the representations of the target user and target item are generated in a sufficiently interactive manner, which can improve the performance of preference prediction as demonstrated by the comprehensive experiments. To the best of our knowledge, the proposed method
is the first attempt towards dynamic or interactive graph convolution for recommendation. In future, we will investigate more complicated attention mechanisms, e.g., extending a single guide to multiple guides or using side information for attention learning.

%since at current we exclusively make use of node features.
% In future work, we will extend 
% In future work, we will follow the direction of attention mechanism and exploit various heterogeneous or homogeneous attention mechanisms considering the node as well as the edge type. On the other side, we will incorporate more side information like item attributes\cite{dong2017hybrid} to assist measurement of attention since at current we exclusively make use of node features.

% \begin{acks}
% \end{acks}

\bibliographystyle{ACM-Reference-Format}
\bibliography{sample-base}

\appendix

\end{document}